\documentstyle[preprint]{aastex}
\received{}
\revised{}
\accepted{}
\journalid{}{}
\articleid{}{}
\paperid{}
\cpright{}{}
\ccc{}

\slugcomment{Submitted to the Astrophysical Journal}

\lefthead{Wood, Whitney, Robitaille, \& Draine}
\righthead{Emission from PAHs \& VSGs in Monte Carlo Radiation Transfer Codes}
\begin{document}

\title{Emission from Very Small Grains and PAH Molecules in 
Monte Carlo Radiation Transfer Codes:\\ 
Application to the Edge-On Disk of Gomez's Hamburger}

%\title{Carcinogens in Gomez's Hamburger}

\author{Kenneth Wood\altaffilmark{1}, Barbara A. Whitney\altaffilmark{2}, Thomas Robitaille\altaffilmark{1}, 
\& Bruce T. Draine\altaffilmark{3} }

\altaffiltext{1}{School of Physics \& Astronomy, University of St Andrews, 
North Haugh, St Andrews, Fife, KY16 9AD, Scotland; kw25@st-andrews.ac.uk}

\altaffiltext{2}{Space Science Institute, 4750 Walnut St. Suite 205, Boulder, CO 80301; 
bwhitney@spacescience.org}

\altaffiltext{3}{Department of Astrophysical Sciences, Princeton University, Peyton Hall, 
Princeton, NJ 08544}

\authoremail{kw25@st-andrews.ac.uk}

\begin{abstract}

We have modeled optical to far infrared images, photometry, and 
spectroscopy of the object known as Gomez's Hamburger. We reproduce the 
images and spectrum with an edge-on disk of mass $0.3M_\odot$ and radius 
1600~AU, surrounding an A0 III star at a distance of 280~pc. Our mass estimate is 
in excellent agreement with recent CO observations. However, our distance 
determination is more than an order of magnitude smaller than previous 
analyses which inaccurately interpreted the optical spectrum. To 
accurately model the infrared spectrum we have extended our Monte Carlo 
radiation transfer codes to include emission from polycyclic aromatic hydrocarbon (PAH) molecules 
and very small grains (VSG). We do this using pre-computed PAH/VSG emissivity files for 
a wide range of values of the mean intensity of the exciting radiation field. 
When Monte Carlo energy packets are absorbed by PAHs/VSGs we reprocess them to 
other wavelengths by sampling 
from the emissivity files, thus simulating the absorption and re-emission process without 
reproducing lengthy computations of statistical equilibrium, excitation and de-excitation 
in the complex many level molecules. 
Using emissivity lookup tables in our Monte Carlo codes gives the flexibility to use 
the latest grain physics calculations of PAH/VSG emissivity and opacity that are being 
continually updated in the light of higher resolution infrared spectra.
We find our approach gives a good representation of the 
observed PAH spectrum from the disk of Gomez's Hamburger. 
Our models also indicate the PAHs/VSGs in the disk have a larger 
scaleheight than larger radiative equilibrium grains, providing evidence for dust coagulation and settling 
to the midplane.

\end{abstract}

\keywords{accretion, accretion disks --- radiative transfer --- stars: circumstellar matter --- 
ISM: dust, extinction}

\section{Introduction}

The object known as Gomez's Hamburger \citep{Ruiz87} exhibits the spectral and imaging 
characteristics of an edge-on circumstellar disk. High resolution HST imagery shows 
a bipolar reflection nebula of size 5.5\arcsec by 2.5\arcsec bisected by an optically thick 
dust lane, reminiscent of 
models and images of disks surrounding low mass Classical T~Tauri stars 
\citep[e.g.,][]{WH92, Burrows96, Grosso2003}. The HST data support the 
circumstellar disk (or torus) interpretation first proposed by \citet{Ruiz87} based on lower 
resolution optical imaging 
polarimetry and the object's broad band spectral energy distribution (SED) which is 
faint in the optical and bright at mid- to far-infrared wavelengths. Such SEDs 
are characteristic of edge-on circumstellar disks \citep[e.g.,][]{Wood2002}. The disk hypothesis 
for Gomez's Hamburger has been strengthened by recent CO observations  \citep{Bujarr2008} 
that are reproduced with a model of a Keplerian disk of mass $0.9 M_\odot (d/500{\rm pc})^2$, 
where $d$ is the distance to the source.

Optical spectroscopy of Gomez's Hamburger indicates the central source is an A0 III star. 
Using parameters typical 
to this spectral type and the integrated broad band spectrum, \citet{Ruiz87}
estimated the object to be at a distance of 2.9~kpc with a circumstellar disk of radius 8000~AU. 
However, as we will show below the disk is smaller and closer than estimated by \citet{Ruiz87}. 
This is because they did not take into consideration that the integrated spectrum 
from an edge-on disk is strongly dependent on inclination \citep{Whitney2003a}. 
At optical wavelengths 
edge-on disks are only seen in scattered light and can be several orders of magnitude 
fainter than systems where the central star is directly seen. In addition to the error in distance determination, 
the spectral type of the central source may not be accurate because the optical spectrum is entirely 
scattered light. Therefore, the combination of high resolution optical spectroscopy 
and radiation transfer modeling of the scattered starlight from the disk is required for more accurate spectral 
typing. However, in the absence of higher resolution optical spectroscopy, in our modeling below 
we adopt the A0 III spectral type from \citet{Ruiz87}.

More recent Spitzer IRS observations of Gomez's Hamburger show strong mid infrared 
emission features typical of emission from polycyclic aromatic hydrocarbon (PAH) 
molecules and very small grains (VSGs). Such emission features are common in massive star 
forming regions \citep[e.g.,][]{Churchwell2006} as the emission mechanism is sensitive to 
the strength of the exciting UV 
radiation from O and B stars. However, charged PAH molecules exhibit significant opacity at 
optical and infrared wavelengths, so that the PAH excitation may arise from softer photons 
\citep{LD2002}, thus enabling PAH emission to be observed from the circumstellar environment of lower 
mass stars. Indeed, PAH emission has been detected from disks and envelopes around A stars 
and classical T~Tauri stars \citep[e.g.,][]{Habart2004, Pontop2007, Geers2006}.

In this paper we present models for the HST images, broad band spectral energy distribution, 
and high resolution Spitzer IRS spectrum of Gomez's Hamburger. In doing so, in \S2 we introduce 
extensions to our Monte Carlo radiation transfer codes that allow us to model 
emission from PAHs and VSGs. \S3 presents the radiation transfer modeling of Gomez's 
Hamburger, and in \S4 we summarize our findings and briefly discuss future applications of our new 
code developments.

\section{PAH/VSG Emission Processes}

Emission from PAH molecules and 
VSGs can dominate the mid-infrared spectrum 
from dusty regions associated with massive star formation.  New 
mid-infrared images from the {\it Spitzer Space Telescope} GLIMPSE survey 
\citep{Churchwell2004, Churchwell2006} dramatically show PAHs acting as a signpost for 
massive star formation.  
The characteristic infrared PAH emission spectrum has been studied in 
detail in many star-forming regions in the Milky Way and other galaxies using a variety of 
ground and space-based observatories 
\citep[e.g.,][]{Peeters2004, DraineSINGS2007}.  
A proper understanding of the 
infrared emission from star-forming regions and galaxies therefore requires 
the inclusion of PAHs and VSGs in radiation transfer codes.

Unlike the radiative equilibrium temperatures attained by large dust grains, 
the absorption and reprocessing of radiation by PAHs and VSGs does not 
yield an equilibrium temperature.  Absorption of a single photon can appreciably increase the energy 
content of the molecule, and the ensuing radiative de-excitation 
produces an infrared spectrum consisting of a continuum plus characteristic emission features.
Detailed modeling of the emission process therefore 
requires solving statistical equilibrium in complex multi-level molecules. 

Approximations exist for determining the reprocessing of photons and emission from PAHs and 
VSGs and have been incorporated into some radiation transfer codes 
\citep[e.g.,][]{Misselt2001, Pontop2007, Dullemon2007, Manske1998}. In this 
paper we introduce a new technique that is readily included into Monte Carlo 
radiation transfer codes and utilizes the detailed PAH and VSG opacity and 
emissivity calculations presented by \cite{DL2007}. 

\subsection{PAH/VSGs in Monte Carlo Radiation Transfer Codes}

Our current Monte Carlo radiation transfer codes \citep[see details in][]{Whitney2003a} 
treat multiple anisotropic scattering in three dimensions and calculate radiative 
equilibrium dust temperatures and spectra using the \citet{BW2001} technique 
or iteratively using methods described by \citet{Lucy1999}. 
The \citet{BW2001} technique works well for two and three dimensional systems \citep[e.g.,][]{Remy2006}, 
but to determine higher signal-to-noise temperatures for the same number of Monte Carlo 
energy packets, especially for three dimensional systems, 
the techniques described by \citet{Lucy1999} are more efficient \citep{CW2008}. 
The \citet{BW2001}
technique produces the same temperatures and spectra as Lucy's method, but requires more 
energy packets (or Monte Carlo photons) and hence longer run times for three dimensional 
simulations. 

\citet{Lucy1999} introduced efficient 
ways for determining mean intensities and hence temperatures and ionization fractions 
in Monte Carlo codes by summing the path lengths within grid cells of individual "energy 
packets" as they take their random walks in the radiation transfer simulation. In particular, 
the Monte Carlo estimator for the mean intensity in a cell is given by,
\begin{equation}
J_i = {\epsilon \over{4\pi\Delta t\Delta V_i}}\sum l
\end{equation}
where the $\epsilon = L \Delta t /N$ is the energy of each of the $N$ Monte Carlo energy packets, 
$L$ is the system luminosity, and $\Delta t$ is the time interval of the Monte Carlo simulation. 
The summation is over all pathlengths, $l$, of energy packets passing through 
cell $i$ of volume $V_i$. We will make use of this Monte Carlo estimator below. 

We have extended our Monte Carlo codes to allow for multiple dust species within any given 
cell in the simulation grid. When Monte Carlo energy packets interact with dust in a cell the 
dust type they interact with is chosen from the relative opacity of each dust species present,
\begin{equation}
P_i = {{\rho_i\kappa_i}\over{\sum_j \rho_j\kappa_j}}
\end{equation}
Where $P_i$ is the probability of the energy packet interacting with dust type $i$ which has 
density and opacity $\rho_i$ and $\kappa_i$.

If a Monte Carlo energy packet is absorbed by a dust type that attains a radiative equilibrium 
temperature, the energy packet is reprocessed and the incremental temperature 
change of the cell is calculated using the Bjorkman \& Wood technique. Therefore at the end of 
the simulation we have a temperature grid for each dust type that attains radiative equilibrium 
temperatures. 

If however, a Monte Carlo energy packet is absorbed by PAH/VSG opacity, the packet is 
reprocessed sampling a new frequency from the pre-computed emissivity files of \citet{DL2007}. 
Since the PAH/VSG emissivity has been computed for a wide range of values of the 
mean intensity of the exciting radiation field, we choose the relevant emissivity file based on 
the value of the mean intensity in the cell where the Monte Carlo energy packet was absorbed. 
This requires an iterative procedure whereby on the first iteration energy packets absorbed by 
PAHs/VSGs are reprocessed using the emissivity file for $J=1$, where $J$ is 
the mean intensity relative to the average value for the interstellar radiation field, 
$J_{ISRF} = 2.17\times 10^{-2} {\rm erg}\, {\rm cm}^{-2}\, {\rm s}^{-1}$ 
\citep{MathisISRF}. On subsequent iterations energy packets are reprocessed 
based on the mean intensity from the previous iteration, calculated using Eq.~1. 
Due to the small PAH/VSG fractions in our simulations (typically less than $4\%$) photons 
reprocessed by PAHs/VSGs do not contribute significantly to the mean intensity. Therefore 
we find that this procedure requires at most three iterations for convergence.

\subsection{Approximations for PAH/VSG Opacities and Emissivities}

The extensions to our Monte Carlo codes of allowing for multiple dust types mean that 
in principle we could employ opacity and emissivity files for grains of many different sizes. 
However, in this first paper we have taken a simpler approach of introducing a cutoff in the 
grain sizes at $a=200$~{\AA}, below which we assume grains are transiently heated and 
energy packets absorbed by these small grains are reprocessed as described above. This 
is a somewhat arbitrary cutoff, but it is a size above which grains generally attain radiative 
equilibrium temperatures \citep{DL2007}.

In addition to introducing a size cutoff between PAH/VSGs and grains in radiative equilibrium, 
we make the assumption that the PAH/VSG emissivity is a function only of the mean intensity 
of the radiation field and not the spectral shape of the exciting radiation field. The \citet{DL2007} 
emissivities that we use were computed for an exciting radiation field with spectral shape of the model 
interstellar radiation field of our Galaxy \citep{MathisISRF}. We make this 
approximation in the absence of emissivity calculations for a wide range of exciting spectra. This is 
an appropriate first approximation since the Monte Carlo code 
accurately treats the transfer and absorption of photons at all wavelengths by the PAH/VSG opacity. 
So in our technique the correct number of photons are reprocessed by PAHs/VSGs, and their 
wavelengths are chosen from the pre-computed mean intensity dependent emissivities. 

We use the opacity and emissivity files for PAHs/VSGs with sizes $a < 200$~{\AA} taken 
from the \citet{DL2007} silicate-graphite-PAH grain model.  Their model assumes 
a power law grain size distribution with the silicate and graphite abundances, PAH charging, and 
PAH mass fraction adjusted to reproduce the opacity and infrared 
spectrum of the Milky Way's diffuse interstellar medium. In particular they find a PAH mass fraction 
of 4.5\% is required to reproduce the observed infrared spectra in the Milky Way and 
other galaxies. 

Figures~1 and 2 show the wavelength dependent opacity and emissivities for PAHs/VSGs 
with grain sizes $a < 200${\AA}. This is the grain size below which we use the 
described approximation for computing the PAH/VSG reprocessing of absorbed energy 
packets. Notice that the PAH/VSG opacity is not negligible at long wavelengths, so 
even photons with just a few eV of energy can excite PAH emission, as described by \citet{LD2002}. 
The shape of the continuum emission in the $20-35\mu$m region is sensitive to the intensity of the illuminating 
radiation field, but the spectral shape and relative strengths 
of the PAH emission features are relatively 
insensitive to U for $U < 10^4$ \citep[see, e.g., Fig. 13 of][]{DL2007}, where $U$ is the energy density 
of the radiation field relative to that of the interstellar radiation field of \citet{MathisISRF}. 
Our reprocessing 
technique accounts for the fact that single photons may excite PAHs/VSGs in environments where 
the mean intensity is very low. Our technique also incorporates the effect of multiphoton excitation 
because the \citet{DL2007} emissivity calculations include this effect which becomes 
important in regions of high mean intensity.

\section{Gomez's Hamburger Models}

Figures 3 and 4 show spectra and images of our model of Gomez's Hamburger that reproduces the 
observed spectrum and {\it HST} images. The broadband optical and near-infrared photometry, as 
well as the IRAS 60\,$\mu$m flux, were taken from \citet{Ruiz87}. The IRAC images and IRS spectrum 
were extracted from the Spitzer Space Telescope data archive (PI: Marwick-Kemper, program P01094). 
Aperture photometry was performed on the IRAC and MIPS 24\,$\mu$m data, with the appropriate point 
source aperture corrections applied. The source is slightly extended at IRAC wavelengths, but SSC 
recommends that for extended sources smaller than 8-9'' in size, the point source corrections should be 
used. The HST WFPC2 images were extracted form the MAST archive (PI: Noll, HST proposal 9315).

The model comprises a flared disk illuminated by a central star.  
The central star was spectroscopically identified as being A0 III by \citet{Ruiz87}, so we adopt  
$L_\star=160 L_\odot$ and $T_\star=10^4$~K. The input spectrum is a $10^4$~K NextGen model 
atmosphere \citep{NextGen}. 
As mentioned earlier, the {\it HST} images resemble those of edge-on protoplanetary disks, 
therefore we adopt a flared disk geometry for the circumstellar density structure parameterized by
\begin{equation}
%\rho (r) =  \rho_0
\rho (r) = \rho_0 \exp\left[{-\frac{1}{2}z^2/h(r)^2}\right]\left( {r\over{R_\star}}\right)^{-\alpha} 
\end{equation}
for $r> R_{\rm min}$ where $r$ and $z$ are the cylindrical coordinates and $R_{\rm min}$ is the 
inner radius of the disk. The disk scaleheight $h(r) = h_0 (r/R_\star)^\beta$ 
where $h_0$ and $\rho_0$ are the scaleheight and density at the stellar radius, $R_\star$. The surface 
density therefore has the form $\Sigma (r) \sim r^{\beta-\alpha}$. Models of disks heated by a central 
star typically yield surface densities $\Sigma \sim r^{-1}$ and flaring parameters in the range 
$1.1\la \beta\la 1.3$ \citep[see review by][]{DullPPV}. 

For the radiative equilibrium circumstellar dust, we adopt the opacity and scattering properties of the 
dust model described in \citet{Wood2002}. This dust model extends to larger sizes than ISM grains and 
produces a shallower wavelength dependent opacity. Larger grains are expected from theories of 
coagulation in disks and this dust model gives a good representation of 
the SEDs of disks in Taurus \citep[e.g.,][]{Rice2003}. We use this dust model in the absence of mm observations 
of Gomez's Hamburger which can help to determine the slope of the long wavelength dust opacity and 
hence information on the dust size distribution.

In addition to a radiative equilibrium dust component we introduce the PAH/VSG 
opacity and emissivity as described above. The PAH/VSG component of the dust has the same 
radial density structure but is allowed to have a larger scale height from the radiative equilibrium 
grains. The scaleheights ($h_0$) of the PAH/VSG and thermal dust components are parameters which are 
varied in a grid of models we ran to reproduce the observations (see below). 
This approximates dust settling in disks \citep[see also][]{Scholz2006} where the smaller 
grains (PAHs/VSGs in our model) have a larger scaleheight than the larger radiative equilibrium 
grains. Therefore the density structures of the PAHs/VSGs and radiative equilibrium grains are 
given by equation~3 with only the $\rho_0$ and $h_0$ terms differing to reflect the different 
mass fractions and scaleheights of radiative equilibrium grains and PAHs/VSGs. 
The mass fraction of PAHs/VSGs is one of the free parameters in our modeling.

In our modeling procedure we ran a small grid of models to determine the distance 
to Gomez's Hamburger in addition to the disk mass, inner hole size, scaleheights of the radiative 
equilibrium dust and PAH/VSG components, and the PAH/VSG fraction required to reproduce the 
features observed in the IRS spectrum. As the disk is observed to be close to edge-on we restricted 
the inclination to be $i > 85^\circ$ because lower inclination disks produce a larger contrast 
between the upper and lower parts of the scattered light disk. 

Assuming the illuminating source is an A0 III star and the circumstellar disk is viewed close to edge-on, 
allows us to determine the distance, interstellar extinction, and disk radius by fitting the optical flux 
levels and image size as shown in figures~3 and 4. From 
this we determine $d = 280$~pc, $R_d = 1600$~AU, and $A_V = 1.8$. 
The disk inclination, flaring parameter ($\beta$),  scaleheight, and mass of the radiative equilibrium dust component are determined from comparing models with the $HST$ 
images. We varied these parameters in the range $85^\circ \le i \le 90^\circ$ and $1.1\le \beta\le 1.3$.

The two SED models shown in figure~3 are our best fit models with and without PAHs/VSGs. 
We calculated goodness of fit $\chi^2$ values for the two models using broadband fluxes only and 
separately using only the IRS data. Assuming 15\% errors on each data point, we find that for the 
broadband fluxes (thirteen data points) the model without 
PAH/VSG effects has $\chi^2 = 381.2$, while the model including PAHs/VSGs has $\chi^2 = 50.8$. Using 
only the IRS data (273 data points) gives $\chi^2 = 19350$ (model without PAHs/VSGs) and $\chi^2 = 376$ 
(model including PAHs/VSGs). For the model including PAH/VSG effects and considering all the data, we 
find $\chi^2 = 1.49$ per data point, i.e., an average deviation of 1.22$\sigma$. 

Many degenerecies exist in modeling disk images and SEDs, notably in the disk flaring parameters 
$h_0$ and $\beta$. Within the context of the smooth, axisymmetric power law disk structure model in 
the parameter space we explored we found values that reproduce the infrared SED and images 
(Figures 3 \& 4) are $i = 87\pm 1^\circ$, $\beta = 1.2$, 
$M_{\rm dust} = 0.003 \pm 0.0005 M_\odot$ 
and $h_0 = 0.0077 \pm 0.0005$. The value of $\beta$ together with our assumption that the 
surface density scales as $\Sigma \sim r^{-1}$  therefore gives $\alpha = 2.2$. Assuming a gas to dust ratio 
of 100 results in a total disk mass of $0.3M_\odot$. This disk mass is in excellent agreement with that 
determined from the CO observations of \citet{Bujarr2008}, who derived a disk mass of 
$0.9 M_\odot (d/500{\rm pc})^2$.

We allow the scaleheight of the 
PAHs/VSGs to differ from that of the larger radiative equilibrium grains and find that a PAH/VSG mass fraction 
of $1.5\pm 0.25$\% and scaleheight $h_0^{PAH} = 0.009\pm 0.001$ reproduces the {\it Spitzer} IRS spectrum as shown in 
figure~3. The larger scaleheight for the PAH/VSG component is consistent with many other disk models that 
predict the coagulation and settling of large grains to the midplane. 
We also found that the IRS spectrum was best reproduced with an inner disk hole size of 
$R_{\rm min} = 500 \pm 25 R_\star$ which is larger than the dust destruction radius of radiative 
equilibrium grains of $40 R_\star$, assuming a typical dust sublimation temperature of 1600~K.

Figures~3 and 4 show that compared to a model without PAHs/VSGs our circumstellar disk model together 
with our new technique for incorporating PAH/VSG emission processes provides a very good match to 
the observed spectral energy distribution 
and high resolution optical images of Gomez's Hamburger. In particular the relative strengths of the 
PAH features in the {\it Spitzer} IRS spectrum are well reproduced, whereas they are absent in the model 
without PAHs/VSGs. We do not fit the IRAS 60$\mu$m data 
point, but this is not a great concern due to potential contamination problems with the IRAS data. In fact the 
IRAS 100$\mu$m data is not useable for this source.

The PAH/VSG emissivities we use in 
our code were developed by \citet{DL2007} specifically to model post-{\it Spitzer} data of the diffuse 
interstellar medium in the Milky Way and other galaxies. 
Further applications of our code to model mid-infrared spectra 
of other disks will enable us to test whether the \citet{DL2007} model can fit PAH features in a broad range of 
environments. Recent studies of PAH features in Herbig Ae stars \citep{Boersma2008} 
suggest there are multiple components that differ from the ISM-like model we have adopted.

In Figure~3 we also show our disk model SED {\it without} PAHs/VSGs which are seen to dominate in the 
wavelength range $5\mu{\rm m}\la\lambda\la 20\mu{\rm m}$ and also with a strong feature around 3$\mu$m. 
The excess emission from disks in this mid-infrared region is very sensitive to disk heating 
by viscous accretion. Therefore care must be taken when analyzing mid-infrared observations of disks 
because radiation transfer models 
that do not include PAH/VSG emission could lead to overestimating the accretion rate in order to ``fill in'' 
the mid-infrared excess. High resolution mid-infrared spectra that can identify PAH features will greatly 
assist in the analysis and interpretation of infrared spectra and accretion rates of protoplanetary disks.

In summary we find that if the central object is an A0 III star, then Gomez's Hamburger lies at a distance of 
280~pc, much closer than the 2.9~kpc estimated by \citet{Ruiz87}. The nebulosity may be modeled with a 
highly inclined disk of mass $0.3M_\odot$ and radius $1600$~AU. 
We reproduce the PAH features present in the {\it Spitzer} IRS spectrum 
using the PAH/VSG model and associated opacities and emissivities computed by \citet{DL2007} for 
their model of the infrared spectrum of the diffuse ISM in our Galaxy.

Our modeling of the circumstellar structure suggests Gomez's Hamburger is an edge-on protoplanetary disk 
around an A0 star. However, if the spectral type of the source is inaccurate, we cannot rule out the possibility 
that Gomez's Hamburger is much closer and the disk has a smaller radius (for a lower luminosity source). Conversely, a higher luminosity central source would yield a larger disk radius and distance. 
As the optical spectrum is entirely scattered light, the combination of high resolution optical spectroscopy 
and radiation transfer modeling of the scattered starlight from the disk is required for more accurate spectral 
typing. The original suggestion from \citet{Ruiz87} is that Gomez's Hamburger is a protoplanetary nebula. 
While our model does not explicitly address this scenario, our derived circumstellar disk structure is consistent 
with dynamical models of planetary nebulae requiring a pre-existing circumstellar disk or torus which can 
act to focus and shape the stellar material ejected during the planetary nebula phase \citep[e.g.,][]{BF2002}. 

\section{Summary}

We have presented a straightforward method to include the absorption and re-emission from PAH molecules and 
VSGs into Monte Carlo radiative equilibrium codes. Our technique utilizes pre-computed intensity-dependent 
emissivities from which we sample the emission spectrum of Monte Carlo energy packets that have been 
absorbed by PAHs/VSGs. Our codes are readily adaptable to include a range of sizes for the 
PAHs/VSGs, but in this first paper we adopt a grain size cutoff of 200~{\AA}, above which we 
assume the dust grains attain radiative equilibrium temperatures. For grain sizes smaller than 200~{\AA}, 
absorbed energy packets are reprocessed using the pre-computed emissivity files as described in \S2. 
Incorporating the PAH/VSG emission in this manner allows us to compute model spectra and images 
very quickly without repeating lengthy calculations of statistical equilibrium and transient heating and 
cooling of the PAH/VSG component --- these calculations are in effect already incorporated in the 
pre-computed PAH/VSG opacities and emissivities presented in figures~1 and 2.

In a future paper we will extend our technique for sampling emissivities to be dependent not only on the 
intensity of the exciting radiation field, but also on the spectrum of the absorbed energy in each grid cell of 
our Monte Carlo simulation. Similar to the mean intensity calculation, determining the spectrum of the 
absorbed energy throughout our grid is straightforward to implement in Monte Carlo radiation transfer 
simulations. This more accurate treatment of the PAH/VSG reprocessing will be possible using planned 
calculations of PAH/VSG emissivity files for a range of exciting spectra comprising model stellar atmosphere 
spectra and black-bodies for a range of temperatures and reddening. 

As a first application of our new code we modeled the object known as Gomez's Hamburger. We find that 
the broad band SED, high resolution images, and {Spitzer} IRS spectrum can be reproduced with an edge-on 
circumstellar disk with a component of PAHs/VSGs comprising 1.5\% of the dust mass.  
Our Monte Carlo radiation transfer using the adopted dust opacities ensures that very few high 
energy photons penetrate into regions of large optical depth. Therefore there is little or no PAH/VSG 
emission from such regions and the PAH/VSG emission is largely confined to a surface layer exposed 
to the direct stellar radiation. This is what is observed in the limb brightened bubbles in the GLIMPSE 
survey \citep{Churchwell2006}. In future papers we will use our code to model the bubbles observed in the 
GLIMPSE survey to address the question of PAH/VSG survival around hot stars and in H~{\sc ii} regions 
in addition to modeling the dust content within the H~{\sc ii} regions and the effects this may have on 
determining the ionizing luminosity of the central sources. We will also investigate emission 
from PAHs/VSGs in protostars, thus extending the radiative equilibrium dust modeling presented by 
\citet{Whitney2003b}.

\begin{figure}
  \plotone{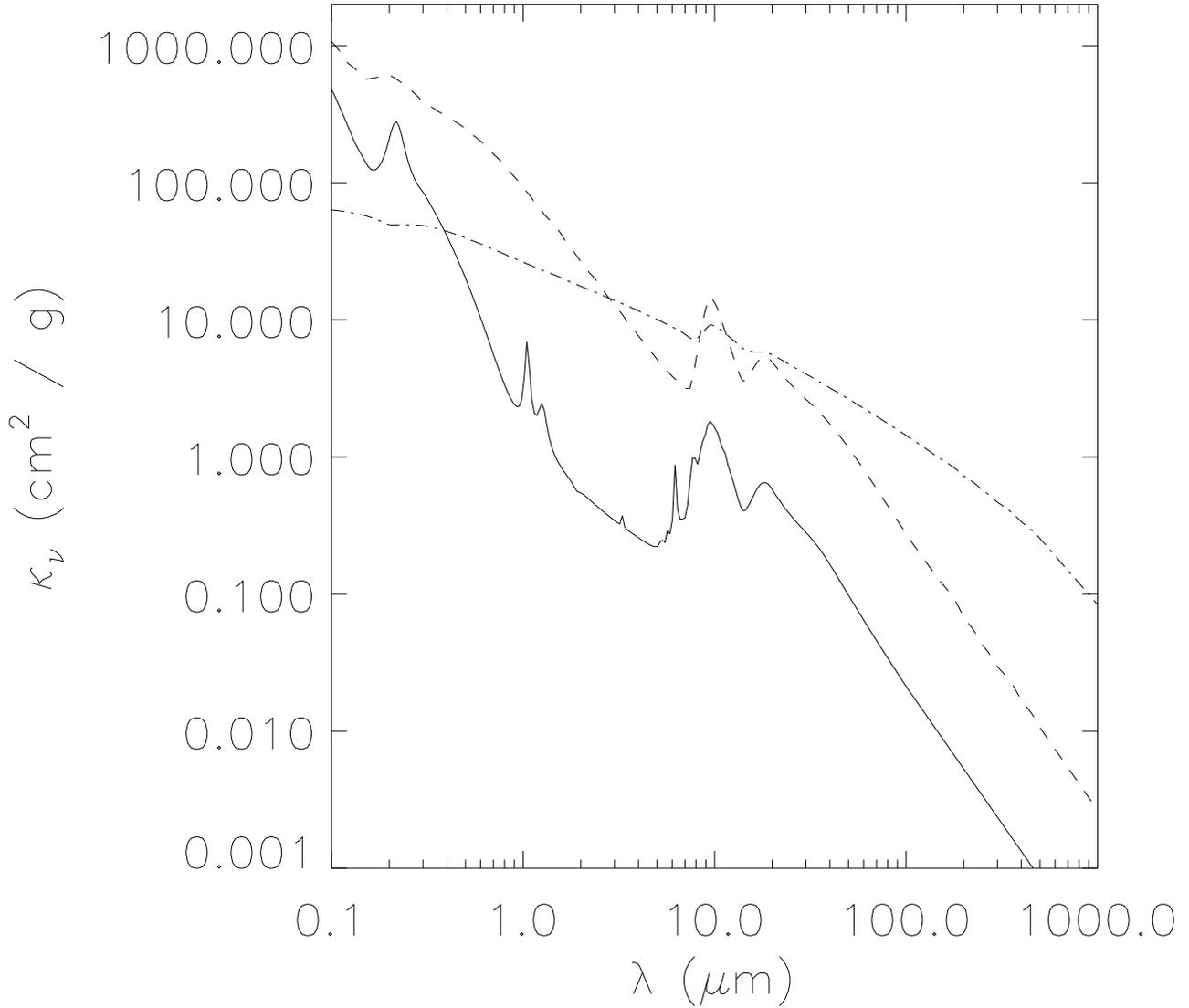}  \caption{Solid line shows the wavelength dependent opacity for the 
  PAH/VSG (dust plus gas) model of 
  \citet{DL2007} for grains with sizes $a < 200$~{\AA}. Dot-dashed line is for the radiative equilibrium 
  dust model of \citet{Wood2002}. For comparison the dashed line shows the opacity for an ISM dust 
  plus gas mixture.}
\end{figure}

\begin{figure}
  \plotone{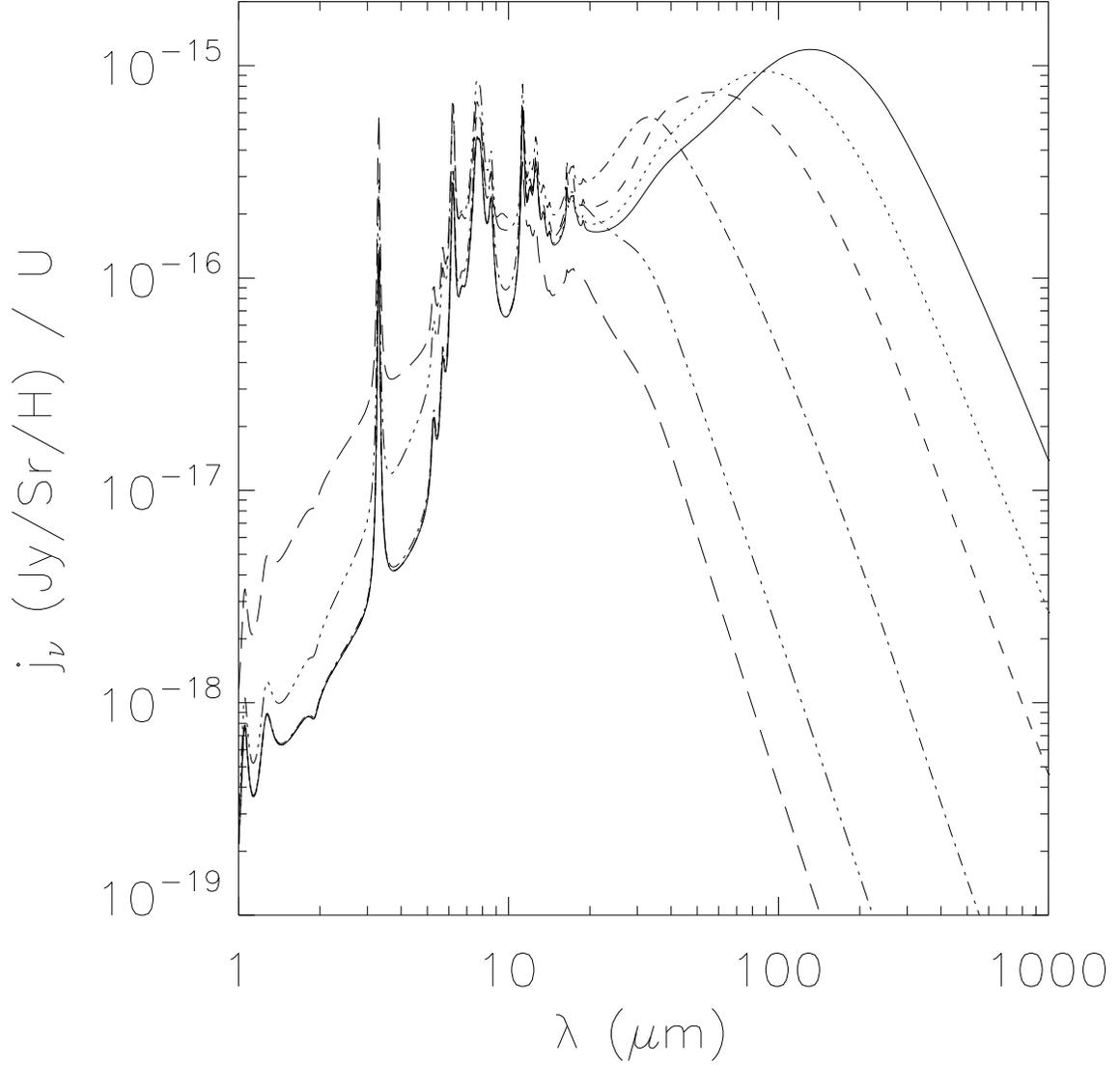}
  \caption{Emissivity for PAHs/VSGs with size $a < 200${\AA} for different values of 
  $U$, the energy density of the exciting radiation field relative to the average interstellar value. The 
  emissivities shown are in the range $1 \le U\le 10^7$. The solid line shows the emissivity 
  for $U=1$ and the long dashed line is for $U=10^7$.}
\end{figure}

\begin{figure}
  \plotone{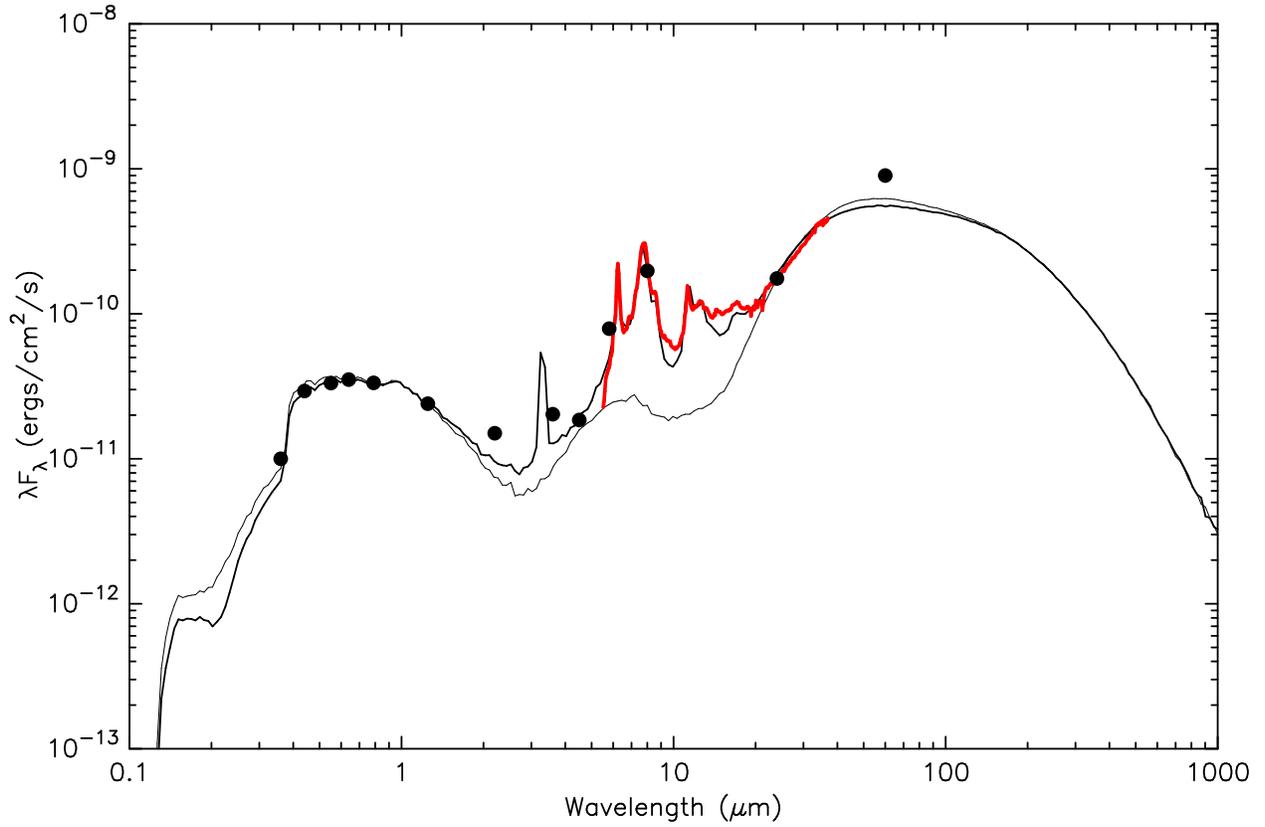}
  \caption{SED data (dots and red curve) and our best disk model (upper solid line) for 
  Gomez's Hamburger. The lower solid line at mid-IR wavelengths 
  shows the same disk model without the PAH/VSG components.}
\end{figure}

\begin{figure}
  \plottwo{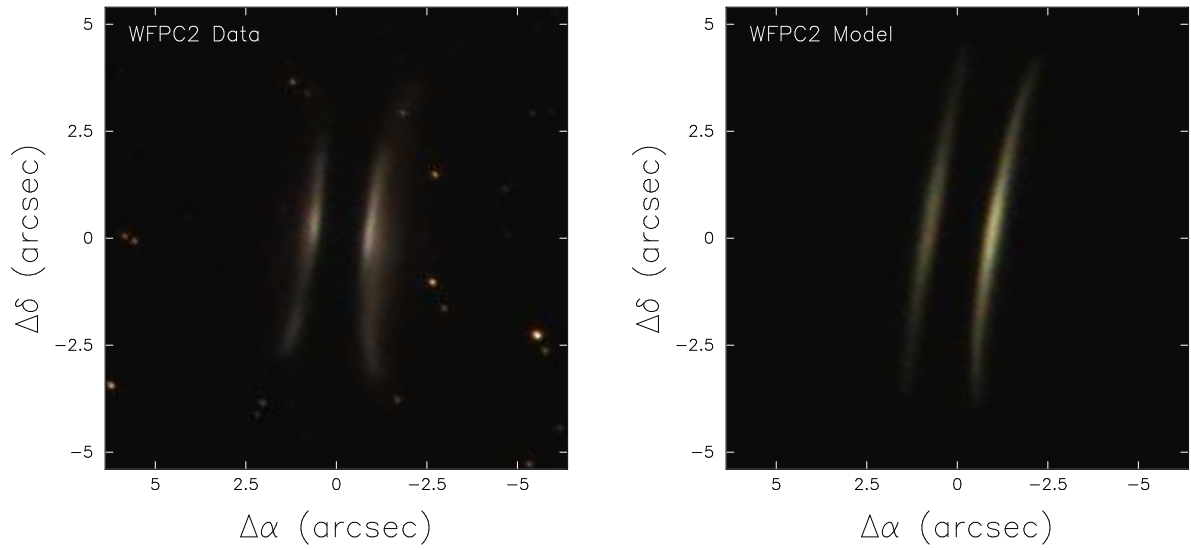}{f4b.eps}
  \caption{Three-color images at $B$ (blue), $V$ (green), and $R$ (red) bands showing HST data (left) and 
  model (right) images for Gomez's Hamburger}
\end{figure}

\acknowledgements

TR is supported by a Scottish Universities Physics Alliance Studentship, 
BAW is supported in part by the NASA Theory Program (NNG05GH35G), 
BTD was supported in part by NSF grant AST-0406883. 

%\bibliography{reference}

\bibliography{ms.bbl}

\end{document}